# Conductivity of DNA probed by conducting-atomic force microscopy: effects of contact electrode, DNA structure, surface interactions

*Thomas Heim, Dominique Deresmes and Dominique Vuillaume[a]*

Institut d'Electronique, Microélectronique et Nanotechnologie - CNRS,

BP69, Avenue Poincaré, F-59652cedex, Villeneuve d'Ascq, France

---

[a] Corresponding author : vuillaume@isen.iemn.univ-lille1.fr

**Abstract**

We studied the electrical conductivity of DNA molecules with conducting atomic force microscopy as a function of the chemical nature of the substrate surfaces, the nature of the electrical contact, and the number of DNA molecules (from a few molecules, to ropes and large fibers containing up to ~ $10^6$ molecules). Independent of the chemical nature of the surface (hydrophobic or hydrophilic, electrically neutral or charged), we find that DNA is highly resistive. From a large number of current-voltage curves measured at several distance along the DNA, we estimate a conductivity of about $10^{-6}$-$10^{-5}$ S·cm$^{-1}$ per DNA molecule. For single DNA molecules, this highly resistive behavior is correlated with its flattened conformation on the surface (reduced thickness, ~0.5-1.5 nm, compared to its nominal value, ~2.4 nm). We find that intercalating an organic semiconductor buffer film between the DNA and the metal electrode improves the reliability of the contact, while direct metal evaporation usually destroys the DNA and prevents any current measurements. After long exposure under vacuum or dry nitrogen, the conductivity strongly decreases, leading to the conclusion that water molecules and ions in the hydration shell of the DNA play a major role.

# I. INTRODUCTION

Molecular electronics has attracted a growing interest owing to its envisioned possibilities to build high-density, low-cost, electronic circuitries. One of the challenging issues deals with the connection of a huge number of molecular-scale devices without the drawback of using traditional electron-beam lithography for the fabrication of electrical wires and contacts. Thus, the demonstration of a highly conducting molecular wire is crucial for future developments. In 1962, Eley and Spivey suggested that π-stacking in double-strand DNA (ds-DNA) could lead to an efficient one-dimensional charge transport.[1] Charge transfer (CT) through DNA molecules was widely studied for a large amount of DNA molecules in solution[2-6] because CT mechanisms have important implications in the damage and repair of this biological system. The conductivity of the "solid-state" thin films of DNA-based compounds was also studied.[7] Recently, DNA molecules deposited on a solid substrate and connected between two electrodes were found highly conducting[8,9,10], insulating[11-14,15,16] or semiconducting.[17,18] These contradictions may come from differences in the base sequence, in the buffer and ambient conditions, in the structural organization of the DNA samples, in the number of DNA molecules in the sample (film, rope, single molecule), in the electrode/DNA coupling, etc. A comprehensive review has recently been published.[19]

In this paper, we report our experiments on electrical conductance in λ-DNA using conducting probe atomic force microscopy (C-AFM). We studied the distance(length)-dependence of the DNA conductance versus the type of the DNA samples: from DNA bundles and ropes to a few single molecules. We also investigated how the transport behavior depends on the chemical nature of the solid surfaces on which DNA molecules are deposited, the nature of the electrical contacts (metallic or organic, electrode directly deposited on the DNA molecules or using large DNA bundles as "buffer").

## II. SAMPLE PREPARATION

### II.1. Surface treatments

We worked with thermally oxidized silicon wafer (300 nm thick oxide grown in dry $0_2$ at 1100°C). We chemically treated the $SiO_2$ surfaces with various silanizating agents: octadecyltrichlorosilane (OTS), oct-7-en-1-trichlorosilane (OETS), and 3-aminopropyltrimethoxysilane (APTMS). Molecules were used as received, and formed self-assembled monolayers (SAM) on the $SiO_2$ surfaces.[20] Alternatively, we also spin-coated a thin film of polystyrene (PS) on the $SiO_2$ surfaces. The above surface treatments allowed us to work with functionalized surfaces exhibiting various surface energies (wettability) and/or electrical charges (see Table I). Extensive wet cleaning (mainly with a piranha solution, $H_2SO_4:H_2O_2$ 2:1, **caution: piranha solution is exothermic and strongly reacts with** **organics)** and dry cleaning by combining ultraviolet irradiation and ozone atmosphere were performed before starting the self-assembly process. Alternatively, we also used an oxygen plasma etching (at 0.1 mTorr, $O_2$ flow of 20 sccm, radio-frequency power 100 W, for 1 min). Both techniques gave a clean, highly hydrophilic, OH-rich surface. The alkylsilane molecules were dissolved ($10^{-3}$-$10^{-1}$ M) in an organic solvent (see details in Table I) maintained at a constant temperature in a dry nitrogen purged glove box. The freshly cleaned substrate was dipped into the solution for times ranging between 30 min and 2 h, depending on the reactivity of the molecules with the surface (Table I). To obtain a densely-packed, well-ordered, monolayer, the Van der Waals (VdW) interactions between the alkyl chains have to overcome the entropic energy. Since the VdW interactions increase with the chain length, and according to the results by Brzoska et al.,[21,22] the smaller the chain is, the weaker the deposition temperature. Thus, we deposited the OTS at 20°C and the OETS at -3°C. One exception is for the amine-terminated molecules (APTMS). These molecules are very short (3 carbon atoms), and it is very difficult to obtain a well-organized monolayer. Most often, these molecules polymerized at the surface forming a film thicker than a monolayer. We also used a deposition technique from a gas phase. Typically, a few ml of APTMS were deposited in a small class flask near (few cm) the sample. Both are put under a glass bell jar filled with dry nitrogen. The sample was allowed to react with the silane vapor for 2 hours. The best results (film thickness and reproducibility) were obtained with this latter process (Table I).

We also treated some of the SiO$_2$ surfaces with a spin-coated polystyrene (PS) film (100-300 nm thick). This PS surface is hydrophobic (see Table I) and was also used for DNA combing.

**II.2 DNA deposition**

The DNA of λ phage (48 502 base pairs, ~16 µm long) was purchased from Roche-Biomedicals. The DNA molecules were dispersed in a TE buffer (10 mM Tris-HCl, 1 mM EDTA) at pH~6-6.5 (Tris = tris(hydroxy-methyl)-aminomethane, EDTA = ethylene diamine tetra acetic acid). This pH value is optimal to comb DNA on various surfaces.[23-25]

**Method A**: A drop (~20-40 µL) of buffer containing DNA (at 250-500 ng/µL, i.e. 10-20 pM, otherwise specified) was deposited on the various surfaces and allowing minutes for incubation. Then, the drop was removed by tilting the sample. The resulting moving liquid-air meniscus allows combing the DNA on the surface perpendicular to the moving meniscus. This technique forms ropes with a small number of DNA per rope (<10). This is a simple variation of the standard DNA combing technique.[23-25]

**Method B**: The drop is deposited on the surface and dried. In that case, bundles and larger ropes are obtained. For characterization by fluorescence microscopy, YOYO$_\circledR$-1 iodile (C$_{49}$H$_{58}$I$_4$N$_6$O$_2$) dyes (Molecular Probes, Eugene, OR) (~ 1 dye every 20 base pairs) were intercalated in the DNA molecules by mixing solutions of both molecules during few hours. Obviously, while YOYO-free DNA molecules were used for AFM and electrical measurements. The fluorescent DNA molecules were only used to assess the quality of the DNA deposition, by measuring the surface concentration, the average DNA length, as a function of the buffer pH and the surface functionality.

**II.3. Contact electrodes**

Conducting-AFM measurements require a "reference" electrode to contact the DNA on one end (Fig. 1), while the other electrode is the conducting-AFM tip. This reference electrode of gold (~10 nm thick) was vacuum evaporated ($10^{-8}$ Torr) through a shadow mask to contact the DNA molecules on one end. In some experiments, we also evaporated an organic semiconductor (pentacene) as the reference electrode. The advantage of using a low melting temperature material (~300-350°C for pentacene vs. ~1000°C for gold) for the evaporated electrodes is to obtain a more reliable contact, avoiding possible destruction of the DNA during the metal evaporation. Obviously, this semiconducting electrode acts as an additional series resistance; this point will be discussed further in section IV. For DNA deposited with method A, some DNA ropes emerged directly from the electrodes and were used for electrical measurements. For DNA deposited by method B, we evaporated the reference electrode on a portion of the DNA "bundle" left on the surface by the drying process (Fig. 1), and measured the currents through the ropes and filaments emerging from this DNA bundle. This latter technique leads to measurable currents (i.e. $>10^{-14}$ A), while all attempts lead to null current when the electrode is directly evaporated on the DNA ropes and filaments, i.e. currents lower than the sensitivity of our apparatus of $10^{-14}$ A (see details in section III).

**II.4. Characterization techniques**

**Contact angle, wetting.** The quality of the monolayers used to functionalize the SiO$_2$ surface was first analyzed by measuring the water and hexadecane contact angles. Water contact angles are measured using deionized water (18 MΩ·cm). We used a remote-computer controlled goniometer system (DIGIDROP by GBX, France) for measuring the contact angles. The accuracy is ± 2°. All measurements were made in ambient atmosphere and at room temperature.

**Ellipsometry.** The monolayer thickness was measured by ellipsometry at 633 nm. In order to estimate the thickness, we used an isotropic value of n=1.50 for the monolayer refractive index at 633 nm and 3.865 for the silicon substrate. Usual values in the literature are in the range 1.45-1.50.[20] One can notice that a change from 1.50 to 1.45 results in an error of less than 1 Å. The accuracy of the monolayer thickness measurements is estimated to be ± 0.2 nm. Alternatively, thickness was also measured my making the SAM into e-beam patterned lines (with widths in the range 1µm to 100 nm) in a polymethylmetacrylate (PMMA) resist.[26] After the lift-off of PMMA (a safe process for the grafted SAM), we measured the SAM thickness by a profil section in TM-AFM image (for details of the process see ref. [26]).

**Fluorescence microscopy.** We used a Leica DMLS microscope with an excitation wavelength of 491 nm and an observation wavelength of 509 nm. The images were recorded with a charge-coupled devive (CCD) camera (Coolsnap-Photometrics).

**Atomic Force Microscopy and Conducting AFM.** We used atomic force microscopy (AFM) to image the surface topography before and after the DNA deposition. A Nanoscope III (Digital Instruments) system in the tapping mode (TM-AFM) was used in air and at room temperature. We used a Conducting-AFM (C-AFM) to locally measure the current-voltage along the DNA. We used a home-made modified Nanoscope III with a PtIr-coated tip. The contact force was controlled by the feedback loop of the Nanoscope, while the current-voltage curve was recorded using an external circuit. The current-voltage (I-V) curve was acquired with an Agilent semiconducteur parameter analyzer HP4145B. The current was first amplified by a home-made current-voltage amplifier (transimpedance gain $G=10^{10}$ $V \cdot A^{-1}$) located nearby the tip and again amplified and filtered by a low-noise voltage amplifier (Standford Research System SR560) before to be recorded by the HP4145B. The detection limit is $10^{-14}$ A. Current larger than 1 nA cannot be measured with this configuration (due to maximum dynamic limit of the current-voltage amplifier). To measure current largers than 1 nA, the home-made current-voltage amplifier can be by-passed and an external current-voltage amplifier (Standford Research System SR590) was used. We recorded the I-V curves at a fixed position along the DNA molecules by applying the tip to the DNA with scanning parameters (x- and y-scans) fixed at zero and at a loading force of 10-30 nN.[27] The C-AFM tip is virtually grounded (input of the transimpedance amplifier) and the voltage is applied on the reference electrode. All measurements were taken at room temperature in ambient air at a relative humidy (RH) of ~50%. It was reported that decreasing RH increases the resistivity of DNA.[28,29] A few of our C-AFM measurements (not reported here) taken under a dry nitrogen flux (RH<20%) confirmed this behavior. Thus, the data reported in this paper concerns the DNA molecules with its hydration layer and counterions. To avoid any excess leakage (tunneling) current between the reference electrode (10 nm thick) and the tip (curvature radius of ~5-20 nm), the minimum electrode-tip distance was set at 50 nm.

## III. RESULTS

**III.1. Characterization of the functionalized SiO$_2$ surfaces**

Table I summarizes the deposition parameters, the water contact angles and the thicknesses of the different organic layers on the SiO$_2$ surfaces. Water and hexadecane contact angles are in agreement with literature data.[20] OTS monolayers show a highly hydrophobic and oleophobic surfaces as expected. The vinyl-terminated monolayer is a less oleophobic as expected by virtue of the presence of the double carbon-carbon bonds. The amine-terminated monolayers are slightly less hydrophobic, with water contact angles that may evolve between 60 and 74°.[30] The OTS and OETS film thickness measurements show a single monolayer since the thickness is in good agreement with the theoretical molecule length (PM3 optimization) assuming that the molecules are in their all-trans conformation as well as standing almost in their up-right position on the surfaces. For the APTMS molecules in solution (toluene), multilayers formed. Their thickness increased (from bilayer to 4-5 layers) with the APTMS concentration in the solution. For the APTMS monolayers formed in gas phase; thickness measurements show that the bilayer formation (table I) was reproducible. The thickness of the PS film is varied from 100 to 300 nm depending on the spin-coater parameters.

**III.2. Characterization of the DNA deposition**

Figure 2 shows typical TM-AFM and fluorescence microscopy images of the DNA deposited by the two methods described in section II.2 on the various treated-$SiO_2$ surfaces. For all hydrophobic surfaces (-$CH_3$, -$CF_3$, -$CH=CH_2$ terminated SAMs and PS films) with method A, we obtained well-aligned ropes of DNA molecules. From the profile section measurements with TM-AFM (taking into account the convolution with the AFM tip shape), we estimated that these ropes contain less than 10 DNA molecules (taking a diameter of ~2.4 nm for the DNA in its B form). On the $NH_2$-terminated surfaces, which are slightly positively charged, DNA combing is more difficult (at pH 6 to 6.5 used in this work) in agreement with previous results.[23] From the analysis of the fluorescence microscope images, we measured the DNA length. The length histograms are shown in Fig. 3. We took into account only ropes with length between 12 µm and 32 µm. Below 12 µm, the DNA molecules are probably broken or are too strongly coiled. Above twice the nominal 16.3 µm length (λ-DNA in B form), several DNA molecules are certainly connected together, and make the sample not suitable for this analysis. In all cases except the amine terminated surfaces, the DNA molecules are overstretched. The peaks in the distribution are at about 27-30 µm for all the hydrophobic surfaces and at ~16-17 µm for the $NH_2$ terminated surfaces. Our values on hydrophobic surfaces are in agreement with other experiments.[23-25] The value for the $NH_2$ surfaces implies that strong interactions between the surface and DNA prevent its stretching by the moving meniscus. Finally, a TM-AFM study on all samples with single (or a few, < ~5) DNA molecules (independent of surface treatment) showed that the height of the DNA molecule is smaller than the expected crystallographic value of 2.4 nm. For the PS, APTMS and OTS treated surfaces, the average heights were 1.17, 1.06 and 1.58 nm, respectively (Fig. 4), with a maximum of samples between 0.5 and 1 nm for PS and APTMS surfaces and between 1 and 1.5 nm for OTS surfaces. This implies that the DNA molecules are distorded and flattened, when deposited onto these surfaces.

With the method B, we obtained bundles and ropes emerging from the bundles (Fig. 2c and 2d). The number of DNA molecules contained in these ropes range from a few up to thousands (as estimated from TM-AFM profile). This result is not strongly dependent on the nature of the surface (hydrophobic, neutral or charged).

**III.3. Conductance**

Figure 5 shows the TM-AFM image of small DNA ropes (deposited by method A on an OTS-treated $SiO_2$ surface) contacted by a gold reference electrode. No measurable current (i.e. $>10^{-14}$ A) was detected by C-AFM for electrode-tip distance larger than 50 nm. The same features were observed for all samples made with deposition method A on all the different functionalized $SiO_2$ surfaces. Replacing Au electrodes by pentacene electrodes led to the same result.

In the case of DNA deposited by method B, and contacted by the reference electrode on the bundles as described in section II.3, two situations were observed (Fig. 6). When the electrode (Au or pentacene) directly contacted the DNA ropes, we detected no current (e.g. through rope #4 in Fig. 6-a). On the contrary, when the contact electrode was evaporated on the bundles (ropes # 1 to 3, Fig. 6-a and Fig. 6-c), we have measured currents through the ropes emerging from this bundle. In that latter case, figure 7 shows several typical I-V curves measured at various distances, d, from the electrode with TM-AFM for various numbers, N, of involved DNA molecules. For the current versus distance measurements along the ropes, the distance d is defined from the boundary of the bundle to the AFM tip (see Fig. 1). The bundle may be considered as a series resistance (see discussion, section IV). A "blank" experiment with the C-AFM tip directly on the chemically treated-$SiO_2$ substrate near the DNA under test showed no measurable current (figure 7-b). All these I-V curves show a common feature. They are asymmetric (figure 7-a) with a stronger current at positive voltage (the voltage was applied on the reference electrode, the C-AFM tip was grounded, see section II). This is consistent with the lower work function for the reference electrode (Au) compared to the tip (Pt/Ir).

In the remainder of this paper, we describe the trends which we oberved over a large number of samples (more than 100) and measurements. These trends of the general behavior and shape of the I-V curves are quite reproducible, althrough we did observe sample-to-sample variations in the quantitative parameters (e.g., the current for a given size and geometry of the sample under test). The I-V curves exhibited a general shape with a slight increase of the current between 0 V and a threshold $V_T$ at which the current increases more rapidly (for instance, $V_T$ is marked on some I-V curves in Fig. 7). When the I-V curve showed a plateau (as curves in Fig. 7-a), we determine $V_T$ as the voltage position of the peak in the first derivative $\delta I/\delta V$ (Fig. 8), otherwise, we estimated $V_T$ by the change in the I-V slope (as curves in Fig. 7-b). We have observed a general tendency (however with some exceptions from sample to sample) that for the small ropes (N<~1000), $V_T$ is between +4 and +7 V and that it decreases for larger systems (Figs. 7 and 8). For very small systems (few DNA molecules), $V_T$ is difficult to estimate due to the very low-level current. Figure 9 summarizes the evolution of $V_T$ versus the estimated number of DNA molecules contained in the ropes. This effect is also visible in Figs. 7 and 8 by comparing the I-V and $\delta I/\delta V$ taken from DNA bundles (very large number of DNA molecules) and the I-V curves taken from ropes (made of about 100 DNA molecules). The resistance is deduced from the first derivative of the I-V curves around a given bias. We have systematically found that the resistance below $V_T$ is about an order of magnitude smaller than above $V_T$. However, for ropes with a small number of DNA molecules, this value was generally below the detection limit. In order to compare the largest amound of data possible; we refer, in the remainder of the paper, to the resistance calculated above $V_T$. All measured electrode/DNA/tip junctions were highly resistive (Fig. 10), from R~$10^9$ Ω for bundles and very large ropes to $10^{15}$ Ω for few DNA molecules. Cai and coworkers measured by C-AFM resistances with the same order of magnitude ($10^9$-$10^{13}$ Ω).[13,14] This also confirms results reported by de Pablo et al.[11] and Storm et al. [12] who showed R>$10^{12}$ Ω for distance larger than few tens of nanometers. Although the same order of magnitude was observed in these various experiments, they are difficult to compare due to possible differences in "series resistances". (i) In our experimental configuration, the bundle between the electrode and the rope (see Fig. 1) acts as an additional series resistance between the rope and the electrode. (ii) Similarly, the organic semiconducting pentacene electrode adds another series resistance as compared to just the gold electrodes resulting in an increase of typically a factor ~10. We found that the advantage of using a low melting temperature material (~300-350°C for pentacene vs. ~1000°C for gold) for the evaporated electrodes is to obtain a more reliable contact, avoiding destruction of the DNA. (iii) The DNA molecules are not covalently attached to the electrodes - we simply have a mechanical contact at the tip end. This adds another series resistance. (iv) The salt in the buffer solution may also crystallize into the DNA bundle during the drying process, and this may affect the conductivity. The TM-AFM images do not show the trace of salt crystallites along the DNA ropes, nor on the surface between ropes. Nevertheless the possible presence of salt in the bundle can be viewed as variations of the series resistance. In some cases, TM-AFM images showed the presence of crystallized salt, however we did not measure I-V curves of these samples. However, all of the above mentioned series resistances have no effect on the relative variation of the currents and resistances versus the distance. Thus we can determine the resistivity per DNA molecule, $\rho_{DNA}$, using the estimated cross-section ($A=NA_{DNA}$ where $A_{DNA}$ is the nominal section of a single DNA molecule ~3nm$^2$) of the ropes and $\rho_{DNA}=A\partial R/\partial d$ (most of our R-d behaviors are nearly linear as shown in Fig. 10). We found that $\rho_{DNA}$ is more or less constant, $\rho_{DNA}$~5x10$^6$ $\Omega$·cm, irrespective of the size of the measured DNA samples and the chemical treatment of the supporting solid surfaces. Fig. 11 shows $\partial R/\partial d$ plotted versus N for our experiments on different chemically treated-SiO$_2$ surfaces. Our $\rho_{DNA}$ value are in agreement with results from De Pablo et al.[11] ($\rho_{DNA}$>10$^6$ $\Omega$·cm),

Storm et al. ($\rho_{DNA}$>~$10^5$ Ω·cm),[12] Okahata et al. ($\rho_{DNA}$~$10^5$ Ω·cm),[7] and Zhang et al. ($\rho_{DNA}$>$10^6$ Ω·cm).[31]

## IV. DISCUSSION

For the small DNA ropes (Fig. 5) deposited by method A, we did not observe any measurable current. This is consistent with the estimated resistivity of $5 \times 10^6$ Ω·cm$^{-1}$ (see previous section). For a single DNA molecules and a distance of 100 nm between the electrode and the C-AFM tip, this corresponds to a current of ~$10^{-15}$ A at 1V, which is below the detection limit of our C-AFM apparatus. In this case, we have observed that the DNA molecules are distorted and flattened, when deposited onto the surface (Fig. 4). This feature could be responsible for the high resistivity reported here. Kasumov et al.[32] recently proposed the same conclusion. They found that a DNA molecule deposited on a pentylamine-treated mica surface has a height of about 2.4 nm and that it is more conducting than DNA on an untreated surface which has a height of ~1 nm. The reason for which the pentylamine film gives the correct DNA thickness is not clear. These authors mentioned that the pentylamine contains $NH_3^+$ molecules. The same is partly true for our amine-terminated surfaces (APTMS, see section III.1) but we did not obtain the correct diameter of DNA in our case (Fig. 4). They also suggested that the pentylamine film decreases the hydrophilicity of the surface, thus decreasing the DNA-surface interactions. This is also the case for our strongly hydrophobic methyl-terminated surfaces (OTS). Fig. 4 shows that DNA molecules on OTS surfaces have the largest diameter (average value of ~ 1.6 nm), which is still lower than the diameter of the native DNA (2.4 nm) and that of DNA deposited on a pentylamine film.[32] More studies are necessary to understand the exact role of pentylamine. The second important geometric factor is the degree of stretching. From fluorescence microscope studies,[33] we found that only DNA molecules on amine-terminated surfaces have the nominal length of 16 µm, while on all the other surfaces DNA molecules are generally overstretched by an average factor of 1.7-1.8 (Fig. 3), in agreement with a previous report.[24] Thus, the fact that DNA is highly resistive, seems to be related to the distorted nature of DNA deposited on solid substrate. Further studies with other treated surfaces, avoiding any distortion of the DNA, is mandatory to draw definitive conclusions. Another possibility is that the deposition of the reference electrode on top of the DNA molecules destroys the DNA underneath, making the electrical contact very resistive. This is supported by the observation that thick DNA ropes allow measurable currents if they are connected to the electrode through a bundle, while no current are detected when the electrode contacts the DNA ropes directly (Fig. 6). The bundle bearing a very large number of interlocked DNA molecules may act as a "buffer" preventing destruction of the DNA submitted to electrode evaporation. Even if some DNA molecules of the bundle are destroyed by evaporation of metals there are still enough DNA molecules intact to establish an electrical contact (while with a high series resistance, as shown when extrapolating data in Fig. 10). The contact resistance, $R_C$, is between $10^{12}$ and $10^{13}$ $\Omega$. According to $R_C = \rho_C L_C / A_C$ with $L_C$ the length of the contact and $A_C$ its cross-section, we can estimate that $\rho_C$ ~ $10^5$-$10^6$ $\Omega \cdot cm$ (from the TM-AFM image of figure 6, $L_C$ is the length of the DNA bundle between the electrode and the onset of the DNA ropes, $L_C$ ~ few µm, and $A_C$ is the product of the bundle width by its thickness, $A_C$ ~ few µm x few hundreds of nm). Thus the contact is as highly resistive as the DNA ropes.

The threshold voltage $V_T$ may be related to the difference between the Fermi energy of the electrodes and the molecular orbitals of the DNA ropes. At $V=V_T$, one molecular orbital of the DNA aligns with the Fermi energy of the electrode, so that carriers can be injected into the DNA molecules. Such resonant band tunneling through DNA has theoretically been calculated, predicting threshold voltages in the range of few volts depending on the energetics and the metal/DNA coupling efficiency.[34,35] Above $V_T$, this current adds to the background current. For the sake of clarity, let us consider a very simplified picture, in which the DNA is sandwiched between the Pt/Ir tip and the gold reference electrode (Fig. 12-a).[36] Under positive bias of the reference electrode, electrons can be injected from the Pt/Ir tip into the LUMO, or holes can be injected from the reference electrode into the HOMO. Depending on the electronic coupling between the electrode and the molecule, only a fraction η (≤1) of the external applied voltage is really applied to the DNA.[37] The first situation would correspond to a weaker coupling between tip and DNA than between DNA and reference electrode (η>0.7 in eq. 1 below, i.e. the major potential drop occurs in the contact barrier with the tip). Conversely, the second injection mechanism, hole injection in the HOMO, should prevail with the opposite coupling scenario (η<0.7). According to the work function of Pt/Ir (~ - 5.4 eV), of Au (~ - 5.1 eV), of the guanine ionization potential, HOMO, (~ - 7.8 eV), while the LUMO's (electron affinities) are higher in energy (~ 0 eV, vacuum level), the corresponding energy barriers are $\Delta_L$ ~ 5.1-5.4 eV and $\Delta_H$ ~ 2.4-2.7 eV, respectively. Thus, the resonance (at positive bias) occurs at[37]

$$V_T^+ = \min\left(\frac{\Delta_L}{e\eta}, \frac{\Delta_H}{e(1-\eta)}\right) = \frac{\Delta_H}{e(1-\eta)} \qquad (1)$$

if η<~0.7 with the energy barriers given above, where η represents the fraction of the potential seen by the molecule (the rest being lost in the contact barrier), and e is the electron charge. In the limit of for η=1 or η=0, $eV_T$ corresponds to the energy barrier. In practice, η<1 depending on the coupling efficiency between the molecule and the electrodes, and thus $eV_T$ is always larger than the expected theoretical values of $\Delta_L$ or $\Delta_H$. The fact that $V_T$ is larger for small ropes (Figs. 8 and 9) actually means that the contact between the C-AFM tip and the DNA ropes is less efficient (higher η) for this ropes than for thick ropes. Similarly, if one molecular orbital comes in resonance at positive bias, we should expect to detect the other one at negative bias. However, due to the smallest current at negative bias (Fig. 7), it was not possible to distinguish any peak in the δI/δV curves at negative voltages (see Fig. 8), the derivatives are too noisy. Also $\Delta_L$ is probably too large to allow electron injection in the DNA molecules. A way to improve the contact is to apply a large voltage (here +8 V) before starting the I-V measurement. A typical example is shown in Fig. 12-b, where we compare the up and down voltage sweep I-V curves. Applying + 8 V for a few seconds before the down voltage sweep leads to higher current and smaller $V_T$ (~1.4 V instead of 5.6 V for the up voltage sweep) as shown by the $\delta I/\delta V$ plots. A better mechanical contact is inferred by the fact that the laser beam detector signal of the AFM cantilever indicates a slightly more pronounced cantilever deflection towards the surface when applying + 8 V before the I-V measurement. A possible explanation for this better contact should be related to an "electrostatic trapping" effect. Using this procedure, $V_T$ is now shifted in the range 1 to 3 V for all sizes of the DNA ropes, in agreement with hole injection in the HOMO through resonant tunneling.[34,35] The fact that $eV_T$ may be lower than the theoratical $\Delta_H$ implies some additional interfacial effects (e.g. charge transfer, interface dipole, etc…) which are to be taken into account. It is also important to note that the resistance at $V \geq V_T$ (that we have discussed here) is not much affected (within a factor of ~ 10). A second general and reproducible effect observed at negative voltages during down voltage sweeps is shown in Fig. 13. For large DNA ropes (> 500 DNA), I-V curves show a similar behavior at negative and positive bias (plateau at $|V|>3V$), while a strong decrease in the current is systematically observed for smaller ropes (<500 DNA) at bias below about –4V. The symmetric behavior is expected if $\eta \sim 0.5$.[37] We surmise that the decrease in current may be due to charge building-up in the DNA rope and space-charge field inhibition of further injections. This is a well-known phenomenon in low carrier mobility materials.[38]

At the beginning of the voltage sweep, charges are initally injected, but since they move slowly, a space-charge build up. Above a critical injected charge level, this space-charge creates a very large internal field opposite to the applied one, which reduces the injected current. This critical charge level depends on many parameters as the carrier mobility, the voltage sweep (i.e. injection time), the DNA rope size (in a smaller rope, the same amount of injected charges results in a larger charge density and thus a larger space-charge field). We can give a crude estimate of this critical amount of injected charges. It takes about 100s to record the I-V curve (from +8V to – 8V) with an average current of 1 pA (Fig. 13). This leads to an estimation of ~$10^9$ injected charges. If we consider ropes with ~1000 DNA molecules (for which we start to see this effect, see Fig. 13), one obtains ~$10^6$ injected charges per DNA. The data in Fig. 13 was recorded for ropes of ~1µm in length. Thus this critical injected charge would correspond to ~$2 \times 10^{23}$ $cm^{-3}$, i.e. a factor ~10 larger than the intrinsic charge of 2 electrons per base pair (~$2 \times 10^{22}$ $cm^{-3}$). Further experiments, varying speed of the voltage sweeps for instance, are mandatory to draw a definitve conclusion. We note that this space-charge effect is never observed for the postive bias. In that case, we inject holes from the reference electrode into the large DNA bundle sitting between the rope and the electrode, while for the negative bias, holes are injected from the C-AFM tip into the DNA rope. The large size of the bundle (see Fig. 6) allows to inject more charges.

These experiments show that there is no significant difference in estimated DNA resistivity regarding the nature of the solid surface mediating the interaction with the DNA, and the stretching of the DNA (see figure 3). Moreover, in both types of experiments a strong increase in the resistivity is observed under vacuum and dry nitrogen. These features support the conclusion that water and counterions in the hydration shell play a main role.[19] Irrespective of the chemically treated surfaces, in an ambient atmosphere, the DNA is always surrounded by its hydration shell. Under long exposures to a dry atmosphere (vacuum or dry $N_2$), the hydration shell is partly removed and the overall conductivity strongly decreases. Another possible effect is that under drying (in nitrogen, RH<10%), the DNA turns from the B to the A form.[39] This structural change may also have an impact on the conductivity of the molecule itself. In the A form, the electronic coupling falls to zero due to the twist angle between the base pairs as predicted by DFT calculations,[19] leading to insulating DNA. However, on the basis of the results presented here, we have no evidence that distinguishes the two effects.

## V. CONCLUSION

We carried out extensive experiments to measure the conductivity of DNA molecules depending on (i) the chemical nature of the solid surfaces on which DNA molecules are deposited, (ii) the nature of the electrical contact (metallic or organic, electrode directly on the DNA molecules or using large DNA bundles as "buffer"), (iii) the number of DNA molecules (from a few molecules, to ropes and large fiber, up to ~ $10^6$ molecules). We conclude that:

(i) in all the explored cases, the electrode/DNA molecules/electrode junction are highly resistive;

(ii) for ropes made of a small number of DNA molecules (≤10) deposited on solid surface (submitted to various chemical surface treatments in order to vary the hydrophobicity/hydrophilicity and/or the charge states), the DNA are distorted and no measurable current can be detected (sensitivity limit of $10^{-14}$ A);

(iii) for ropes made of a larger number of DNA molecules (≥10) deposited on a solid surface, non linear current-voltage curves are measured by conducting-AFM. From the current vs. distance behavior a conductivity of $10^{-6}$-$10^{-5}$ S·cm$^{-1}$ per DNA is deduced;

(iv) reproducible currents can be obtained only when the DNA ropes are connected to the evaporated electrode through a large DNA bundle, which probably act as a "buffer" to prevent an extensive defect creation under vacuum metal evaporation;

(v) water molecules and counterions in the hydration shell around the DNA play a significant role. After a long exposure under vacuum or dry nitrogen, the conductivity strongly decreases.

**Acknowledgements**. We thank for discussions H. Bouchiat, A.Y. Kasumov and V. Croquette. Part of this work was financially supported by IRCICA (Institut de Recherche sur les Composants pour l'Informatique et la Communication Avancées). One of us (DV) would like to acknowledge stimulating discussions with F. Rondelez on chemical surface treatments and surface interactions with DNA molecules.

**Table I**: Deposition condition, contact angle and thickness of the organic films (self-assembled monolayers or spin-coated films) used to chemically treat the silicon dioxide surface prior to the DNA deposition.

| Surface treatment (molecule) | Chain length (number of C atoms) | Solvent | Concentration (mM) | Temperature of deposition (°C) | Deposition time | Contact angle (water/hexadecane) (±2°) | Thickness[b] (± 0.2 nm) |
|---|---|---|---|---|---|---|---|
| OTS | 18 | HD/CCl$_4$ (60:40) | 1-5 | 20 | 2 h | 108/43 | 2.5 (2.54) |
| OETS | 8 | HD/CCl$_4$ (60:40) | 1-5 | -3 | 2 h | 99/(<10)[f] | 1.3-1.5 (1.3) |
| APTMS | 3 | Toluene | 1-5 | 20 | 30 min | 60-74/(<10)[f] | 1.7 to 3.9 (0.86)[d] |
| APTMS | 3 | (a) | (a) | 20 | 2 h | 60-74/(<10)[f] | 1.4-1.7 (0.86) |
| Polystyrene | na | Toluene | (c) | 20 | (c) | 90/(<10)[f] | 100-300 nm[e] |

(a) : gas phase deposition
(b) : in brackets, the theoretical molecule length (PM3 optimization) in its all trans conformation
(c) : spin-coating in toluene
(d) : increasing with concentration
(e) : depending on spin-coating parameters (speed, etc) and concentration
(f) : contact angles lower than 10° cannot be accurately measured with our setup.
HD = hexadecane
na = not applicable

# Figure captions

**Figure 1** : Scheme of the experimental set-up showing two typical situations : DNA ropes and filaments connected to the electrode through a DNA bundle (top) and DNA ropes or filaments directly connected by the electrode (bottom).

**Figure 2** : (a) TM-AFM images (3 µm x 3 µm) of DNA deposited by method A on an OTS-treated surface. (b) Fluorescence microscope image (64 µm x 45 µm) of DNA deposited by method A on a $NH_2$-terminated surface (APTMS). (c) TM-AFM images (8 µm x 8 µm) of DNA deposited by method B on an OTS-treated surface. (d) TM-AFM images (6 µm x 6 µm) of DNA deposited by method B on a $NH_2$-terminated surface (APTMS).

**Figure 3** : Histograms of the DNA length measured from fluorescence microscope images for DNA deposited by method A on OTS, OETS, APTMS and PS surfaces.

**Figure 4** : Histograms of the DNA height measured by TM-AFM for single, isolated, molecules (or at least a few molecules) deposited on various treated surfaces (PS, APTMS - i.e. $NH_2$ terminated and OTS – i.e. $CH_3$ terminated).

**Figure 5 :** TM-AFM image of a small DNA ropes (< 5 DNA molecules) deposited by method A on OTS-treated surface and connected by an evaporated gold electrode (white area).

**Figure 6 :** (a) TM-AFM image (8 µm x 8 µm) of DNA bundles and ropes deposited by method B on amine-treated surface (APTMS) and contacted by a pentacene layer (above the marked line on the image). In C-AFM, ropes #1, 2 and 3, with a DNA bundle "buffer" between the rope and the contact, give a measurable current, while rope #4, directly brought into contact with the electrode, does not. (b) TM-AFM image (6 µm x 6 µm) of a network of DNA ropes (deposition method B on an APTMS surface). The electrode (not shown) is evaporated on the bundle seen at the top of the image. (c) Corresponding C-AFM image (z-scale is 400 fA) taken at an applied tip bias of 5V.

**Figure 7 :** Typical current-voltage (I-V) curves. (a) Large DNA ropes (~1200 to 3700 DNA) deposited by method B on APTMS surface. 3 I-V are taken with the C-AFM tip at ~300nm from the boundary of the DNA bundle (■), at ~600nm (●) and at ~2 µm (▲). The number of DNA decreases while increasing the distance, ~ 3700 at 300nm, ~2000 at 600nm and ~1200 at 2µm. (b) Small DNA ropes (~100 molecules) deposited by method B on PS surface. 4 I-V curves are shown : (■) directly on the DNA bundle, (●) at ~300 nm from the boundary of the bundle, (▲) at ~600nm and (▼) directly on the surface (blank experiment) nearby the DNA ropes in test. On some of these I-V curves, the threshold $V_T$ is marked by an arrow.

**Figure 8 :** First derivative $\delta I/\delta V$ for I-V curves: (a) $\delta I/\delta V$ for a small rope (~500 molecules) displaying a larger $V_T$ at ~5.1 V, (b) corresponds to (▲) of Fig. 7-a, (c) to (●) of Fig. 7-a and (c) to (■) of Fig. 7-a. The threshold voltages $V_T$ (peaks in the $\delta I/\delta V$ shown by a vertical line) are almost the same (2.5-2.8 V) for these DNA ropes with more than 1000 DNA molecules.

**Figure 9 :** Variation of the threshold voltage $V_T$ as a function of the size of the DNA system (estimated number of DNA from cross-sections of TM-AFM images).

**Figure 10 :** Plot of the resistance (measured from I-V curves above $V_T$) versus distance for a large number of samples with various sizes of the DNA sample: (●) ~1000, (▼) ~600, (▲) ~350, (■) ~300, (◆) ~220 DNA molecules. DNA molecules were deposited by method B on PS-treated surfaces.

**Figure 11 :** Linear resistance (δR/δd) versus the estimated number of DNA in the ropes and for DNA deposited on differently treated surfaces.

**Figure 12 :** (a) Simple energy level diagram of the electrode/DNA/C-AFM tip junction. (b) I-V and δI/δV curves for a rope of about 500 DNA for the up and down voltage sweeps. Before the down voltage sweep, a + 8V bias was applied for few seconds.

**Figure 13 :** I-V curves (down voltage sweep) at negative bias for a rope of ~500 DNA molecules (○) and a rope of ~1000 DNA molecules (□).

Figure 1, T. Heim et al.

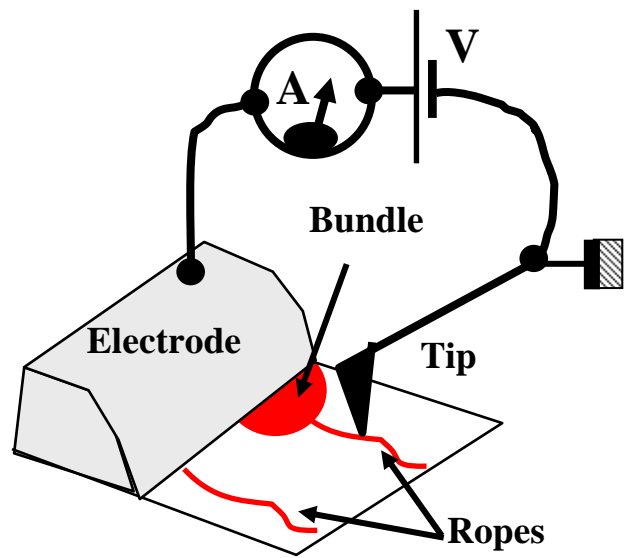

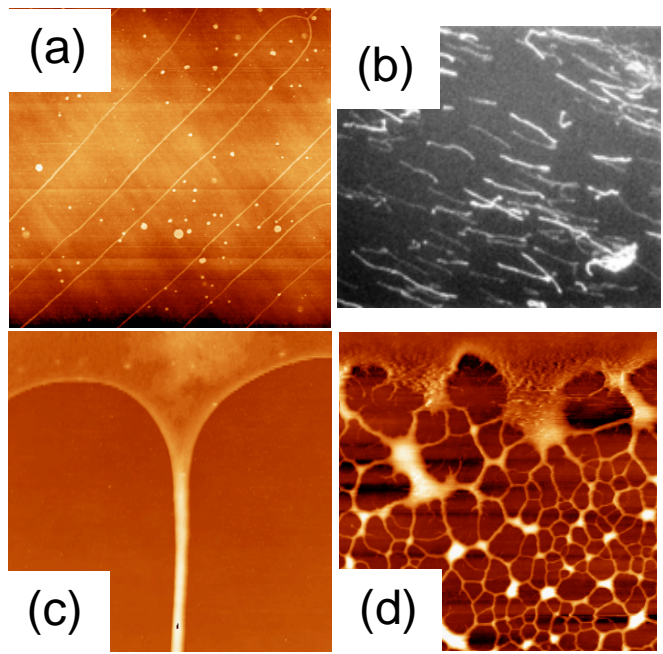

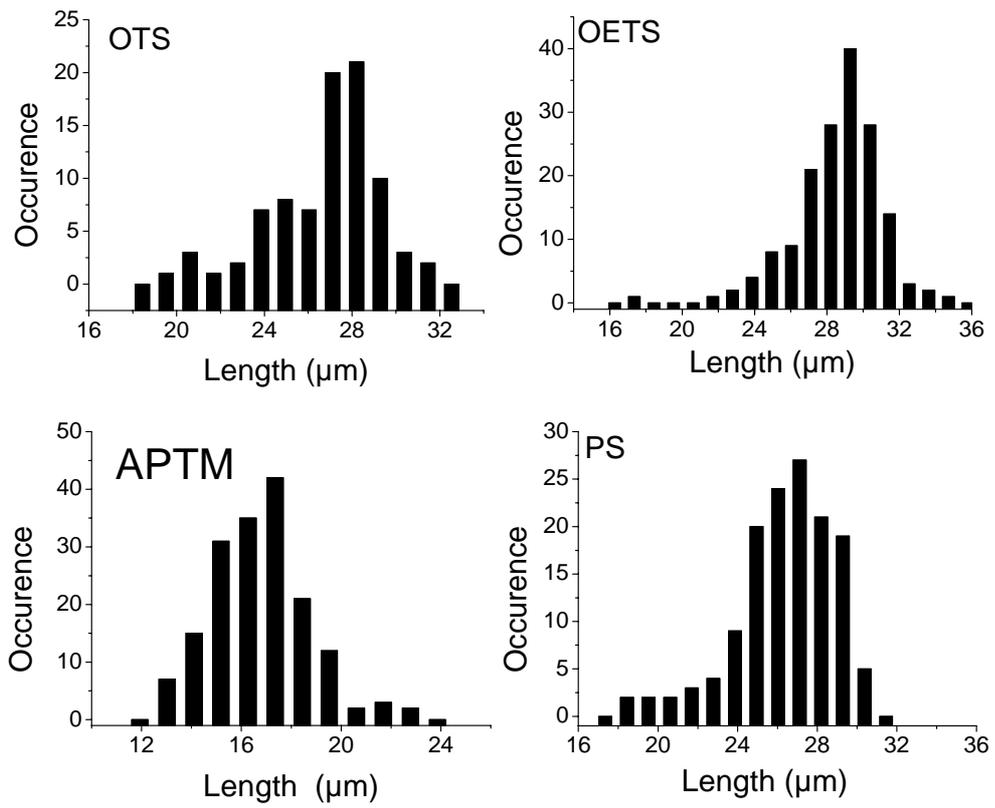

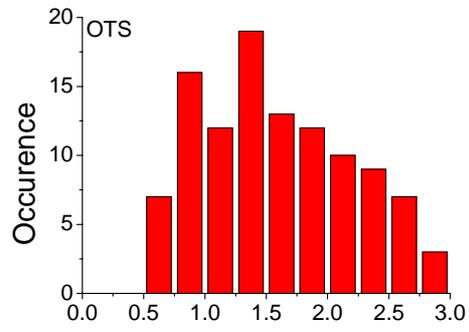
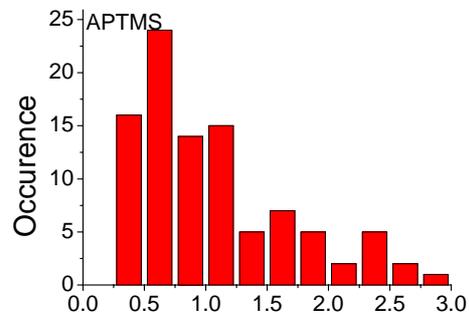
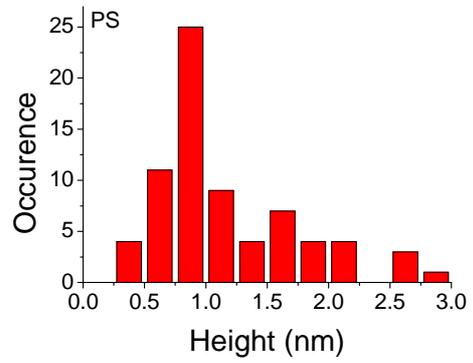

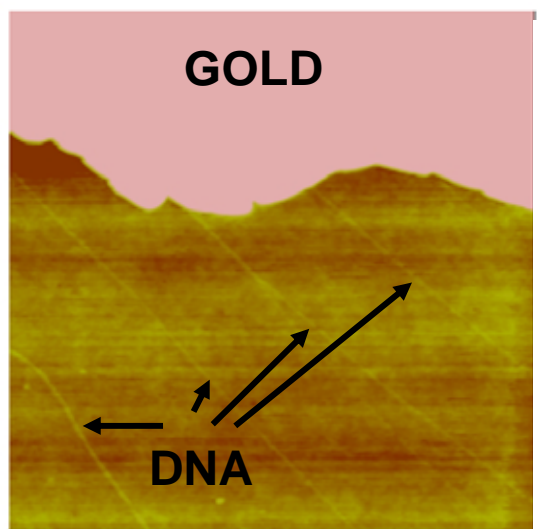

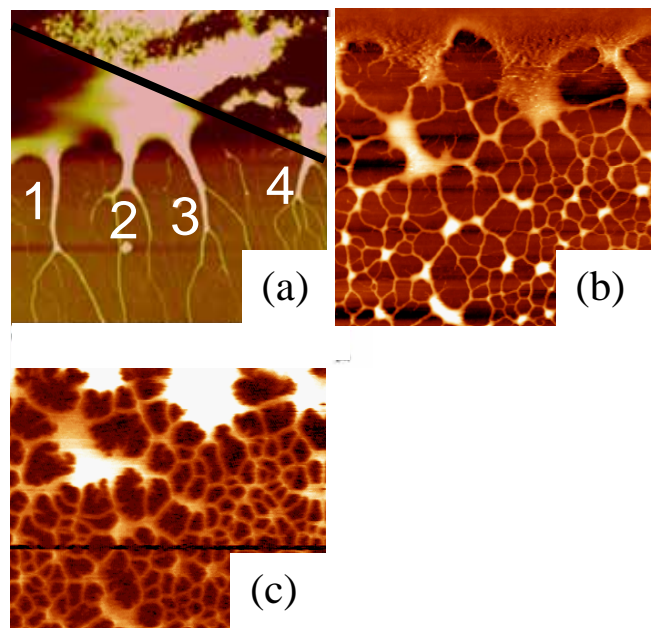

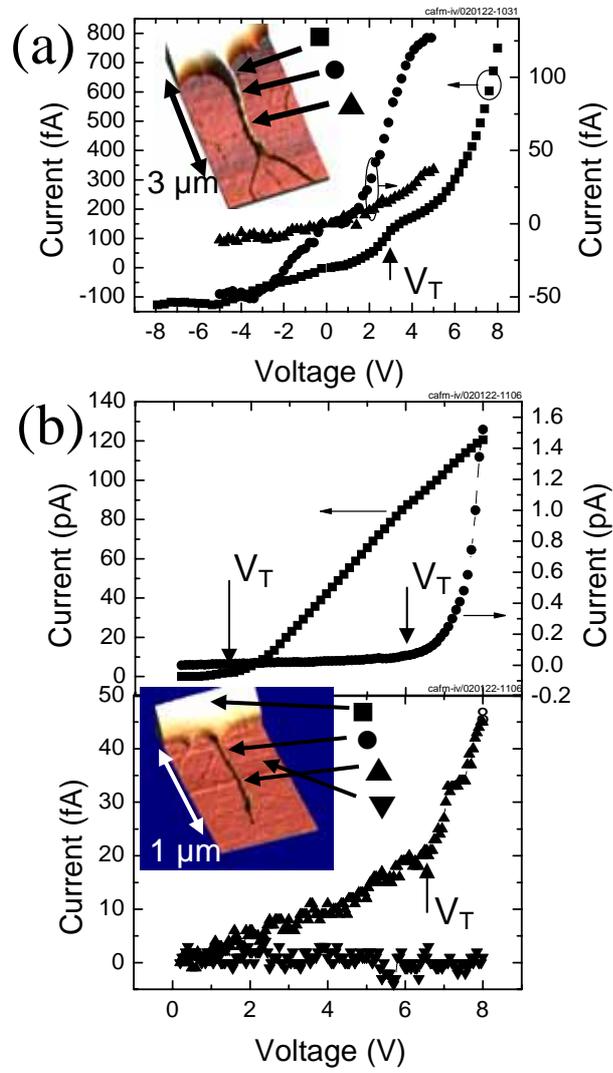

Figure 8, T. Heim et al.

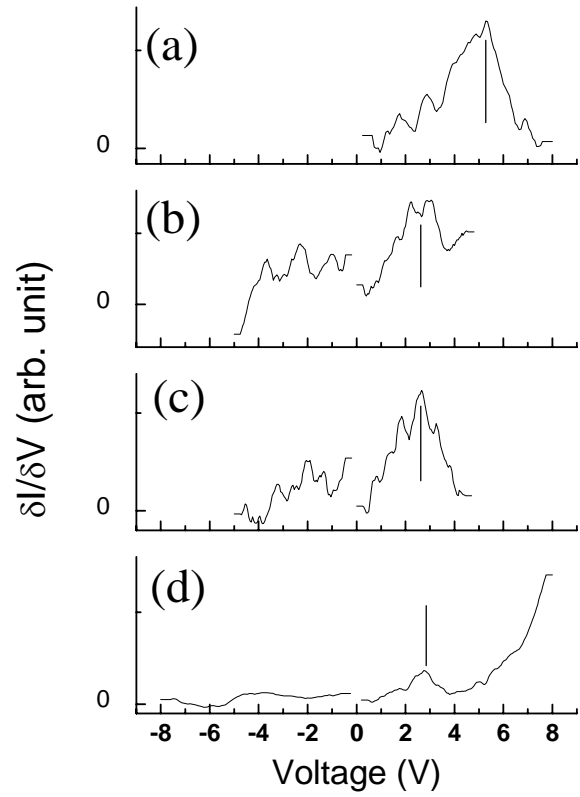

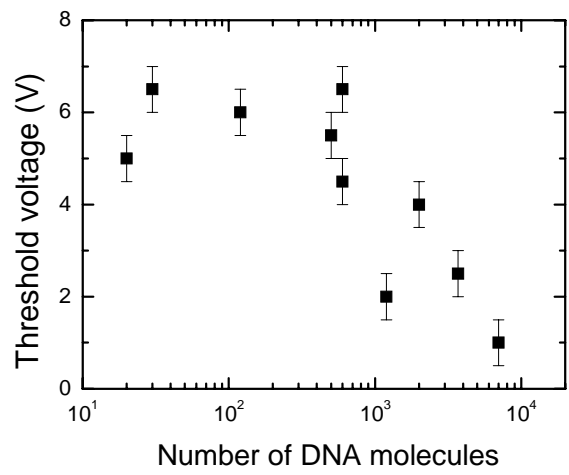

Figure 10, T. Heim et al.

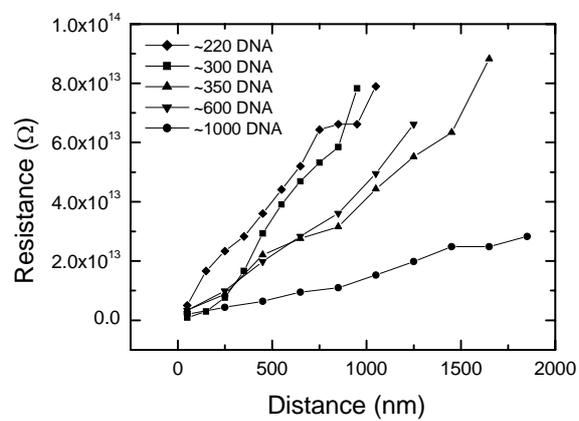

Figure 11, T. Heim et al.

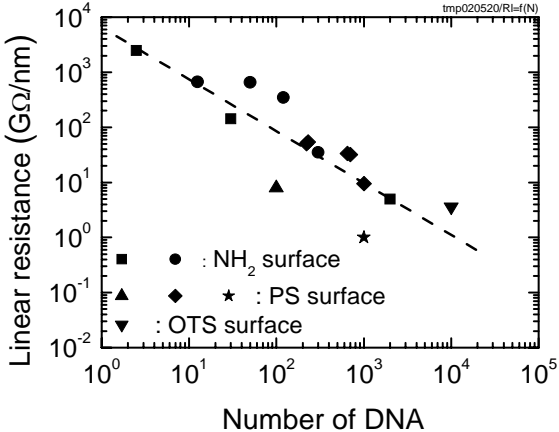

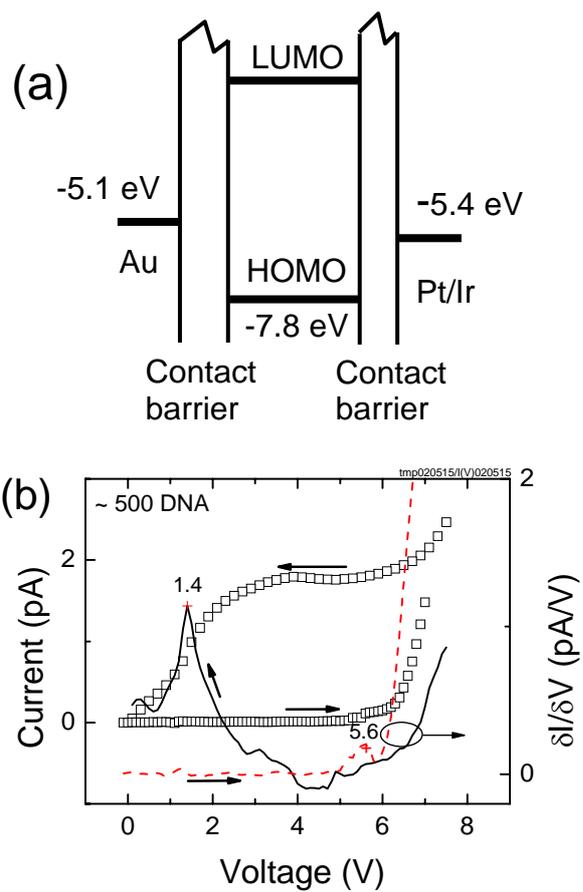

Figure 13, T. Heim et al.

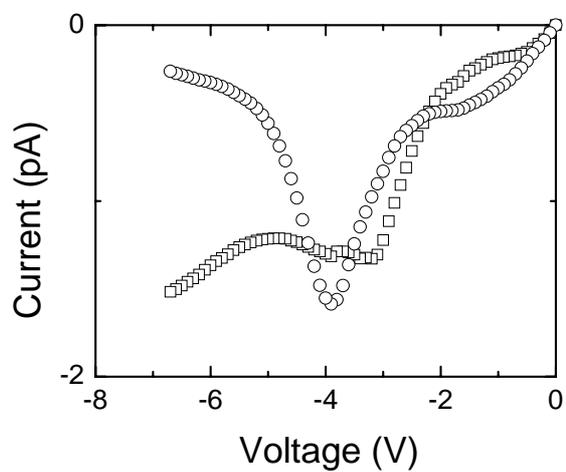